\documentclass[11pt]{article}
\begin{document}
\title{\bf Background field quantization and non-commutative Maxwell theory}
\author{Ashok Das$^{\rm a}$, J. Frenkel$^{\rm b}$,
    S. H. Pereira$^{\rm b}$ and J. C. Taylor$^{\rm c}$\\
\\
$^{\rm a}$ Department of Physics and Astronomy,\\
University of Rochester,\\
Rochester, NY 14627-0171\\
USA\\
\\
$^{\rm b}$ Instituto de F\'{\i}sica,\\
Universidade de S\~ao Paulo,\\
S\~ao Paulo, SP 05315-970, BRAZIL\\
\\
$^{\rm c}$ Department of Applied Mathematics\\ 
and Theoretical Physics,\\
University of Cambridge,\\
Cambridge, UK}

\date{}
\maketitle

\begin{center}
{ \bf Abstract}
\end{center}

We quantize non-commutative Maxwell theory canonically in the background field
gauge for weak and slowly varying background fields. We determine the 
complete basis for expansion under such an approximation. As an
application, we derive the Wigner function which determines the
leading order high temperature behavior of
the perturbative amplitudes of non-commutative Maxwell theory. To
leading order, we also give a closed form expression for the distribution
function for the non-commutative $U (1)$ gauge theory at high temperature.

\newpage

\section{Introduction}

Background field techniques are quite useful in the study of
conventional non-Abelian gauge theories
\cite{honerkamp,background}. In particular, they simplify
calculations enormously, since in a background field gauge, invariance
under background gauge transformations is manifest. On the other hand,
canonical quantization of non-Abelian gauge theories within such a
framework is highly
nontrivial \cite{ambjorn}. The difficulty comes from the fact that a
complete  basis for the field equation of the quantum gauge field is
not easy to determine in general.  The
only known example of a successful canonical quantization is for the
case where the background field strength is a constant \cite{ambjorn}.
However, there are many physical situations where the background field
can be considered to be weak and the variations in
the background field less rapid than the variations in the quantum
field. In such a case, we show that canonical quantization in the
background field gauge can be carried out and, in fact, we demonstrate
this within the context of non-commutative Maxwell theory. (The
quantization goes through unchanged even in the presence of fermion
fields.) 

The high temperature behavior of a plasma \cite{books,das,
  braaten, frenkel} constitutes an
example where the external field is assumed to be weak with slow
variations compared with the quantum fields. Namely, in the hard
thermal loop approximation,  it is normally assumed that
\begin{equation}
p\ll k \sim T\, ,
\end{equation} 
where $p$ represents a typical external momentum while $k$ denotes an
internal loop momentum. Our method finds a natural application in the
study of such systems and we derive the Wigner function for
non-commutative photons which, in turn, determines 
the leading high temperature behavior of amplitudes in this theory. In
an earlier work \cite{brandt}, we had studied this question
through  the use of Wigner function (without the use of the background
field method), where we had noted some peculiarity of noncovariance
of the results under a gauge transformation and had argued for a
covariant completion of a particular form. In the present approach,
we show 
that the covariance of the results is manifest and that the particular
covariantization found earlier, results from the proper quantization in
the background field gauge. This, therefore, clarifies
the meaning of the covariant completion found in \cite{brandt}. When working
within the framework of background field method, an alternate and
simpler definition of Wigner function (to the leading order) is
possible. Using this, we are also able to determine the leading order
distribution function in a closed form at one loop.

The organization of our results is as follows. In section {\bf 2}, we
briefly recapitulate the basics of the background field
method. Considering the leading order behavior of the equation of
motion as well as 
the background gauge condition, in section {\bf 3} we determine a
complete basis for the covariant D'Alembertian operator
in this approximation which allows for an expansion of the quantum gauge
field. As an application of our method, in section {\bf 4} we introduce an
alternate, simpler definition of the Wigner
function which describes the leading order behavior of the
amplitudes through a transport equation and show that the calculations
are manifestly covariant and agree with the perturbative
results. Furthermore, using this Wigner function, we determine a
closed form expression for the leading order distribution function at
the one loop level. 

\section{Background field method in non-commutative Maxwell theory}

In this section, we review very briefly the basics of background field
method within the context of non-commutative Maxwell theory. For our
conventions, 
notations and the definitions of star product etc, we refer the
reader to \cite{brandt} as well as the vast literature on the subject of
non-commutative  field theories \cite{douglas, szabo}. 

Non-commutative $U (1)$ gauge theory (Maxwell theory) is described by
the action 
\begin{equation}
S [A] = \int \mathrm{d}^{4}x\, \left(-\frac{1}{4} F_{\mu\nu}\star
F^{\mu\nu}\right)\, ,\label{qedaction}
\end{equation}
where
\begin{equation}
F_{\mu\nu} = \partial_{\mu} A_{\nu} - \partial_{\nu} A_{\mu} - ie
\left[A_{\mu} , A_{\nu}\right]\, .
\end{equation}
Here, the commutator stands for the Moyal bracket of the fields. We
now make the expansion
\begin{equation}
A_{\mu} = \bar{A}_{\mu} + a_{\mu},\quad \langle a_{\mu}\rangle = 0\,
,\label{decomposition} 
\end{equation}
where $\bar{A}_{\mu}$ is the background field satisfying
\begin{equation}
D_{\mu} (\bar{A}) F^{\mu\nu} (\bar{A}) = \bar{D}_{\mu}
\bar{F}^{\mu\nu} = \partial_{\mu} \bar{F}^{\mu\nu} - ie
\left[\bar{A}_{\mu} , \bar{F}^{\mu\nu}\right] = 0\, ,\label{equation}
\end{equation}
$a_{\mu}$ is the quantum field and $\langle a_{\mu}\rangle$  denotes
the expectation value of the quantum field in any given state. Then, it
follows that 
\begin{equation}
F_{\mu\nu} = \bar{F}_{\mu\nu} + \bar{D}_{\mu} a_{\nu} - \bar{D}_{\nu}
a_{\mu} - ie \left[a_{\mu} , a_{\nu}\right]\, .
\end{equation}
In such a case, the action (\ref{qedaction}) can be expanded as
\begin{eqnarray}
S [\bar{A}+a] & = & \int \mathrm{d}^{4}x\,\left[-\frac{1}{4}
  \bar{F}_{\mu\nu}\star \bar{F}^{\mu\nu} - \frac{1}{2}
  (\bar{D}^{\mu}a^{\nu} - \bar{D}^{\nu}a^{\mu})\star \bar{D}_{\mu}
  a_{\nu} + ie \bar{F}^{\mu\nu}\star a_{\mu}\star
  a_{\nu}\right.\nonumber\\
 &  & \qquad\qquad \left. + ie (\bar{D}^{\mu}a^{\nu} -
  \bar{D}^{\nu}a^{\mu})\star a_{\mu}\star a_{\nu} + \frac{e^{2}}{2}
  \left[a^{\mu} , a^{\nu}\right]\star a_{\mu}\star a_{\nu}\right]\, ,
\end{eqnarray}
where linear terms in the quantum field do not occur by virtue of the
equations of motion (\ref{equation}) for the background field.

The advantage of the background field method lies in the fact that the
original gauge invariance of the theory under
\begin{equation}
\delta A_{\mu} (x) = \partial_{\mu}\epsilon (x) - ie \left[A_{\mu} (x) ,
  \epsilon (x)\right]\, ,
\end{equation}
can be viewed in one of two ways. First, this can be thought of as a
quantum gauge invariance under the transformations,
\begin{eqnarray}
\delta \bar{A}_{\mu} (x) & = & 0\, ,\nonumber\\
\delta a_{\mu} (x) & = & \partial_{\mu}\epsilon (x) - ie
\left[\bar{A}_{\mu} (x) + a_{\mu} (x) , \epsilon
  (x)\right]\, ,\label{quantum} 
\end{eqnarray}
or as a background gauge invariance under
\begin{eqnarray}
\delta \bar{A}_{\mu} (x) & = & \partial_{\mu} \epsilon (x) - ie
\left[\bar{A}_{\mu} (x) , \epsilon (x)\right]\, ,\nonumber\\
\delta a_{\mu} (x) & = & - ie \left[a_{\mu} (x) , \epsilon
  (x)\right]\, .\label{background}
\end{eqnarray}
Namely, under a quantum gauge transformation, the background field is
inert while the quantum field transforms like a gauge field. On the
other hand, under a background gauge transformation, the background
field transforms like a gauge field while the quantum field transforms
in the adjoint representation. It is, therefore, possible to take
advantage of this and add the gauge fixing and the ghost actions
\begin{equation}
S_{\rm GF} + S_{\rm ghost} = \int \mathrm{d}^{4}x\,\left[- \frac{1}{2\xi}
  (\bar{D}\cdot a)\star (\bar{D}\cdot a) +
  \bar{D}^{\mu}\bar{c}\star (\partial_{\mu}c - ie \left[\bar{A}_{\mu}
  + a_{\mu} , c\right])\right]\, ,
\end{equation}
which break the quantum gauge invariance (\ref{quantum}), but are
invariant under the
background gauge transformation (\ref{background}) with ghosts
transforming in the adjoint representation. As a
result, calculations carried out with such a background gauge fixing
will lead to results that are manifestly invariant under background
gauge transformations.

The part of the total action quadratic in the quantum fields is
responsible for the one loop results and has the form
\begin{eqnarray}
S_{q} & = & \int \mathrm{d}^{4}x\left[-\frac{1}{2} (\bar{D}^{\mu} a^{\nu} -
  \bar{D}^{\nu} a^{\mu})\star \bar{D}_{\mu} a_{\nu} + ie
  \bar{F}^{\mu\nu}\star a_{\mu}\star a_{\nu}\right.\nonumber\\
 &  & \quad\qquad\left. - \frac{1}{2\xi}
  (\bar{D}\cdot a)\star (\bar{D}\cdot a) + \bar{D}^{\mu}\bar{c}\star
  \bar{D}_{\mu}c\right]\, .\label{quadratic}
\end{eqnarray}
Here $\xi$ represents the gauge fixing parameter and in the limit
$\xi\rightarrow 0$ we have the background field gauge condition,
\begin{equation}
\bar{D}\cdot a = \bar{D}_{\mu}a^{\mu} = \partial_{\mu}a^{\mu} - ie
\left[\bar{A}_{\mu} , a^{\mu}\right] = 0\, .\label{gauge}
\end{equation}
In this gauge, the equation of motion for the quantum field (at the one
loop level) follows from (\ref{quadratic}) to be
\begin{equation}
\bar{D}_{\nu} \bar{D}^{\nu} a^{\mu} = 2ie \left[\bar{F}^{\mu\nu} ,
  a_{\nu}\right]\, .\label{equation1}
\end{equation}
It can be easily checked that (\ref{gauge}) and (\ref{equation1}) are
compatible.

\section{Quantization}

To quantize the gauge field, we have to determine a basis satisfying
both (\ref{gauge}) and (\ref{equation1}). This is, in general, a very
hard problem and a solution in a factorizable form may not always
exist.  We,
therefore, look for a solution of these equations in the approximation
that the background fields are weak and slowly varying compared to the
quantum fields. As we have pointed out in the introduction, there are
various physical phenomena of interest that satisfy such conditions and
in  the
next section we will have an application to such a physical
situation. With these assumptions, Eq. (\ref{equation1}) reduces in
the leading order to
\begin{equation}
\bar{D}^{2} a^{\mu} = \bar{D}_{\nu}\bar{D}^{\nu} a^{\mu} =
  0\, .\label{equation2}
\end{equation}
Thus, in this approximation, it is essential to determine a basis for
the covariant D'Alembertian operator in order to quantize the gauge
field. Furthermore, since $a_{\mu}$ transforms covariantly under a
background gauge transformation (\ref{background}), the basis function
must reflect this. We know that the plane waves $e^{ik\cdot x}$
represent a basis for the D'Alembertian operator in the absence of any
background gauge fields. Correspondingly, let us denote the basis in
the presence of background gauge fields as $e_{\star}^{ik\cdot X}$.

To determine this basis, let us define (see also \cite{brandt})
\begin{equation}
\tilde{A}_{\mu} = \bar{A}_{\mu} + \frac{1}{k\cdot
  \bar{D}}\,\bar{F}_{\mu\nu} k^{\nu} = \frac{1}{k\cdot
  \bar{D}}\,\partial_{\mu} (k\cdot \bar{A})\, .\label{tilde}
\end{equation}
It is clear that this transforms like a background gauge field under
(\ref{background}). We note from (\ref{tilde}) that, in general,
\begin{equation}
k\cdot \tilde{A} = k\cdot \bar{A} \, .
\end{equation}
Furthermore, we see from (\ref{tilde}) that in the gauge
\begin{equation}
k\cdot \bar{A} = 0\, ,
\end{equation}
$\tilde{A}_{\mu}$ vanishes so that it must be a pure gauge field
satisfying 
\begin{equation}
k\cdot \tilde{A} = \frac{i}{e} \Omega^{-1}\star k\cdot\partial
\Omega\, .\label{puregauge}
\end{equation}
This can also be equivalently written as
\begin{eqnarray}
k\cdot \partial\Omega + ie \Omega\star k\cdot \tilde{A} & = & 0\,
,\nonumber\\
k\cdot\partial \Omega^{-1} - ie k\cdot \tilde{A}\star \Omega^{-1} & = & 0\,
.\label{covariant}
\end{eqnarray}

The solution of (\ref{puregauge}) (or equivalently of
(\ref{covariant})) can be determined easily in terms of link operators
and has the form
\begin{equation}
\Omega (x) = U^{(\tilde{A})} (-\infty,x) = {\rm
  P}\left(e_{\star}^{-ie\int_{-\infty}^{0}
  \mathrm{d}u\,k\cdot\tilde{A} (x+ku)}\right)\, ,\label{solution}
\end{equation}
where the integration is along a straight path parallel to
$k^{\mu}$  from $-\infty$ to $x^{\mu}$. For simplicity,
we have taken the reference point to be at $-\infty$ although this is
not  necessary. We note here that open Wilson lines of the form in
(\ref{solution}) play an important role in non-commutative gauge theories
from various points of view at zero temperature
\cite{douglas,szabo,raamsdonk} (including in the 
construction of gauge invariant observables \cite{wilson}) as well as
at finite temperature in the construction of effective actions
\cite{brandt1}. From the form of the solution
in (\ref{solution}) as well as the properties of star products, it
immediately follows that
\begin{equation}
\Omega^{-1} (x) \star e^{ik\cdot x}\star \Omega (x) = U^{(\tilde{A})}
(x,-\infty)\star U^{(\tilde{A})} (-\infty, x+ \theta k)\star e^{ik\cdot x}
= U^{(\tilde{A})} (x, x+\theta k)\star e^{ik\cdot x}\,
,\label{composition} 
\end{equation}
where we have introduced the notation
\begin{equation}
(\theta k)^{\mu} = \theta^{\mu\nu} k_{\nu}\, ,
\end{equation}
and used the property
\begin{equation}
e^{ik\cdot x}\star f (x) \star e^{-ik\cdot x} = f (x+\theta k)\, .
\end{equation}
From the transformation properties of the link operators, we see that
the  combination of factors on the left hand
side of (\ref{composition}) transforms covariantly under
(\ref{background}). 

From the definition in (\ref{puregauge}) (or (\ref{covariant})) as
well as the fact that $k^{2} = 0$, it follows that
\begin{equation}
k\cdot \bar{D}\left(\Omega^{-1} (x)\star e^{ik\cdot x}\star \Omega
(x)\right) = 0 = k\cdot \bar{D}\left(U^{(\tilde{A})} (x, x+\theta
k)\star e^{ik\cdot x}\right)\, .\label{vanishing}
\end{equation}
With this, we now see that, to leading order in our approximation,
\begin{equation}
\bar{D}^{2} \left(U^{(\tilde{A})} (x, x+\theta k)\star e^{ik\cdot
  x}\right) =  ik\cdot \bar{D}\left(U^{(\tilde{A})} (x, x+\theta
  k)\star e^{ik\cdot x}\right) = 0\, ,
\end{equation}
where we have neglected the term where the covariant derivative acts
on the link operator since the momentum of the background field is
sub-leading compared to $k$. Thus, we see that to leading order in our
approximation, a basis for the covariant D'Alembertian operator can be
written as
\begin{equation}
f_{k} (x) = e_{\star}^{ik\cdot X} = U^{(\tilde{A})} (x, x+\theta
k)\star e^{ik\cdot x}\, .\label{basis}
\end{equation}
This transforms covariantly under (\ref{background}) and reduces to
ordinary plane waves when $\bar{A}=0$ (and, therefore,
$\tilde{A}=0$ or when $\theta^{\mu\nu}=0$). We also note that following the
derivation  in \cite{raamsdonk}
(where an integrated form of the relation is obtained), it
is easy to show that 
\begin{equation}
U^{(\tilde{A})} (x,x+\theta k)\star e^{ik\cdot x} = e_{\star}^{i(k.x +
  e k\times \tilde{A} (x))}\, ,\label{covariantcoordinate}
\end{equation}
where we have used the notation standard in non-commutative field
theories,
\begin{equation}
A\times B = \theta^{\mu\nu} A_{\mu}B_{\nu}\, .
\end{equation}
Equation (\ref{covariantcoordinate}) represents precisely the covariantization
factor determined earlier in \cite{brandt} from different
considerations and  this
derivation clarifies the origin of such a factor, showing that we can
think  of $X^{\mu}$ (see (\ref{basis})) as the appropriate
covariant coordinate for the problem at hand.

The basis for the covariant D'Alembertian operator allows us to make
an expansion of a covariant (from the non-commutative $U(1)$ point of
view) scalar field. However, the expansion of a gauge field must, in
addition, 
satisfy (\ref{gauge}). To that end, we note that in the leading order
of our approximation, Eq. (\ref{gauge}) can be written as
\begin{equation}
\partial_{\mu} a^{\mu} (x) = 0\, ,\label{gauge1}
\end{equation}
which is the Landau gauge. Correspondingly, expansion in terms of the
usual transverse polarization vectors is sufficient to satisfy the
gauge condition in the leading order. We would like to emphasize that,
in general, the polarization vector can have sub-leading terms that are
not necessarily factorizable (which has been checked to lowest orders) which
is another reason why quantization in the background field method  is
highly  nontrivial in general. With all this, we can now expand the
quantum fields (in the leading order of our approximation) as
\begin{equation}
a_{\mu} (x) = \sum_{s} \int
\frac{\mathrm{d}^{3}\bar{k}}{(2\pi)^{3}\sqrt{2\bar{k}^{0}}}\,\epsilon_{\mu}
(\bar{k},s) 
\left(a (\bar{k},s) e_{\star}^{-i\bar{k}\cdot X} + a^{\dagger}
(\bar{k},s) e_{\star}^{i\bar{k}\cdot X}\right)\, ,\label{expansion}
\end{equation}
where we have assumed without loss of generality that the polarization
vector is real and
\begin{equation}
\bar{k}^{0} = |\bar{\bf k}|,\quad \bar{k}\cdot \epsilon (\bar{k},s)
= 0\, .\label{conditions} 
\end{equation}
The quantum field can now be quantized and the physical Hilbert space
identified in the standard manner \cite{kugo}.

\section{Application}

As an application of the background field quantization of the previous
section, we will now derive the Wigner function \cite{elze} which
determines  the leading order high temperature
behavior of the amplitudes in non-commutative QED. We note that
conventionally the covariant Wigner function for the non-commutative
photon is defined as (see \cite{brandt})
\begin{equation}
W_{\mu\nu} (x,k) = \int \frac{\mathrm{d}^{4}y}{(2\pi)^{4}}\,e^{-iy\cdot k}\,
G_{\mu\lambda}^{(+)} (x)\star G^{\lambda (-)}_{\;\nu} (x)\,
,\label{wigner} 
\end{equation}
where
\begin{equation}
G_{\mu\nu}^{(\pm)} (x) = U^{(A)} (x, x_{\pm})\star F_{\mu\nu}
(x_{\pm})\star U^{(A)} (x_{\pm},x),\quad x_{\pm} = x\pm \frac{y}{2}\, .
\end{equation}
On the other hand, within the framework of the background field method
an alternate and simpler definition of the covariant Wigner function
\cite{elze1},
which describes the leading order behavior in the hard thermal loop
approximation, is possible and has the form 
\begin{equation}
w_{\mu\nu} (x,k) = \int \frac{\mathrm{d}^{4}y}{(2\pi)^{4}}\,e^{-iy\cdot k}\,
G_{\mu}^{(+)} (x) \star G_{\nu}^{(-)} (x)\, ,\label{newwigner}
\end{equation}
\begin{equation}
G_{\mu}^{(\pm)} (x) = U^{(\bar{A})} (x,x_{\pm})\star a_{\mu}
(x_{\pm})\star U^{(\bar{A})} (x_{\pm},x)\, .\label{gmu}
\end{equation}
There are several things to note from (\ref{newwigner}) and
(\ref{gmu}). First, the Wigner function in (\ref{newwigner})
transforms covariantly under (\ref{background}) independent of whether
the link operator in (\ref{gmu}) is defined with respect to the
complete gauge field
$A_{\mu}$ or with respect to the background gauge field
$\bar{A}_{\mu}$. However, we have defined it with respect to the
background field to avoid some problems that arise otherwise in
a practical calculation. (Such problems also arise in the conventional
definition and need various assumptions on the factorizability of
thermal correlation functions. However, a definition such as in
(\ref{gmu})  avoids such
assumptions.) Since the Wigner function (\ref{newwigner}) is already quadratic
in the quantum fields, at one loop level, the gauge fields in the link
operators would factor out of the thermal correlation functions as
background  fields even if we use the complete gauge field to define
the link operators. However, keeping an eye on the potential
difficulties that may arise at higher loop level from such terms (if
defined with a complete gauge field), we have chosen the particular
definition in (\ref{gmu}). Second, the Wigner function in
(\ref{newwigner}) can be easily seen to be related to the one in
(\ref{wigner}) in the leading hard thermal loop approximation as
\begin{equation}
\eta^{\mu\nu}\langle w_{\mu\nu} (x,k)\rangle = -
\lim_{k^{2}\rightarrow 0} 
\frac{\eta^{\mu\nu}}{2k^{2}}\,\left(\langle W_{\mu\nu} (x,k)\rangle -
\bar{W}_{\mu\nu} (x,k)\right)\, ,\label{identification}
\end{equation}
where $\langle..\rangle$ denotes thermal average.

Following the derivation in \cite{brandt}, the transport equation for
$w_{\mu\nu}$ can now be derived. In fact, if we define the
distribution function 
\begin{equation}
{\cal F} (x,k) = \eta^{\mu\nu} \langle w_{\mu\nu} (x,k)\rangle \,
,\label{distribution}
\end{equation}
then, it can be easily shown that (this also follows from 
equation (33) derived in \cite{brandt} and the identification in
(\ref{identification}))
\begin{equation}
k\cdot \bar{D} {\cal F} (x,k) = \frac{e}{2} \frac{\partial}{\partial
  k_{\sigma}} k^{\rho} \left[\bar{F}_{\rho\sigma}\star {\cal F} +
  {\cal F}\star \bar{F}_{\rho\sigma} - 2 \int
  \frac{\mathrm{d}^{4}y}{(2\pi)^{4}}\,e^{-iy\cdot k}\,\langle
  G_{\mu}^{(+)}\star \bar{F}_{\rho\sigma}\star G^{\mu
  (-)}\rangle\right]\, .\label{transport}
\end{equation}
By iteratively solving the transport equation (\ref{transport}), the
distribution function (\ref{distribution}) can be determined to any order in
the leading approximation. This would, then,  determine the current
defined as  
\begin{equation}
J_{\mu} (x) = - e \int \mathrm{d}^{4}k\,\theta (k^{0}) \left\{k_{\mu}
\left({\cal F} (x,k) - {\cal F} (x,-k)\right)\right\}\, ,\label{current}
\end{equation}
which, in turn, would yield the leading order amplitudes of the
theory through functional differentiation.(By leading order in this
context, we refer to the class of terms inside the curly bracket in
(\ref{current}) 
which, apart from a $\delta (k^{2})$, are functions of
zero degree in $k$. In conventional QCD, this class yields all the
dominant contributions, but in noncommutative QED, this may not give
the complete contribution for arbitrary values of the noncommutative
parameter $\theta^{\mu\nu}$ as discussed in \cite{brandt1}.) The
anti-symmetrization in
the definition in (\ref{current}) is to ensure the correct charge
conjugation property \cite{sheikh-jabbari} of the current as discussed
in \cite{brandt}.

With the field expansion given in (\ref{expansion}), the thermal
averages can be calculated in the physical space \cite{das} satisfying
\begin{equation}
\sum_{s=1,2} \epsilon_{\mu} (\bar{k},s) \epsilon^{\mu} (\bar{k},s) = - 2\, ,
\end{equation}
and the current can be determined order by order. It is
straightforward to check that this coincides with the results obtained in
\cite{brandt} (which
also gives the appropriate perturbative amplitudes). However, unlike
the earlier work, here the results are manifestly covariant at any
order with the expansion in (\ref{expansion}), without any need for
covariantization. Furthermore, in this case, the distribution function
in (\ref{distribution}) can even be obtained in a closed form in the
leading order at one loop and has the form
\begin{eqnarray}
{\cal F} (x,k) & = & 2 \int \frac{\mathrm{d}^{4}y}{(2\pi)^{4}}\,e^{-iy\cdot
  k} \int \frac{\mathrm{d}^{4}\bar{k}}{(2\pi)^{3}}\,\delta
  (\bar{k}^{2})n_{B} (|\bar{k}^{0}|)\nonumber\\
 &  & \qquad\times \left\{\theta(\bar{k}^{0})\left(e_{\star}^{\frac{y\cdot
  \bar{D}}{2}} 
  e_{\star}^{i\bar{k}\cdot X}\right)\star \left(e_{\star}^{-\frac{y\cdot
  \bar{D}}{2}} e_{\star}^{-i\bar{k}\cdot X}\right)+ (\bar{k}\leftrightarrow
  -\bar{k})\right\}\, ,\label{distribution1}
\end{eqnarray}
where $n_{B} (|\bar{k}^{0}|)$ denotes the Bose-Einstein distribution and the
covariant translation (for a covariant function under the
non-commutative $U(1)$) is explicitly given by
\begin{equation}
e_{\star}^{\pm \frac{y\cdot \bar{D}}{2}} f (x) =
U^{(\bar{A})}(x,x_{\pm})\star f (x_{\pm})\star U^{(\bar{A})}(x_{\pm},x)\, .
\end{equation}
Using the transport equation (\ref{transport}), the distribution
function in (\ref{distribution1}) can be systematically expanded order
by order in the number of background fields (or powers of $e$) and
substituted  into the
definition of the current in (\ref{current}). We have verified
explicitly, up to the three photon amplitude,  
that this reproduces correctly the perturbative results in the leading
order in the hard thermal loop approximation.

\section{Conclusion}

In this work, we have quantized non-commutative $U (1)$ gauge theory
canonically in the background
field method using the background field gauge for weak and slowly
varying background fields. We have determined a (covariant) basis for the
covariant D'Alembertian operator which indeed coincides with the
particular covariantization factor determined earlier from different
considerations. We have applied our quantization method to study the
high temperature behavior of non-commutative Maxwell theory in the
leading  order
using the Wigner function. The calculations are manifestly covariant
and agree with the perturbative results. We have also determined a
closed form expression for the distribution function for the photon in
the leading approximation at one loop. Although our discussion has
been within the context of non-commutative Maxwell theory, this can be
generalized to conventional QCD as well. In particular, a basis for
the covariant D'Alembertian operator can again be constructed in terms
of a pure gauge (background) field $\tilde{A}_{\mu}$ defined in
(\ref{tilde}). This can then be used to determine, in principle, the
leading order distribution function in QCD. This is an interesting
question that deserves further study.

\vskip .5cm

One of us (JF) would like to thank F. T. Brandt for helpful
discussions. 
This work was supported in part by US DOE Grant number DE-FG
02-91ER40685 as well as by CNPq, FAPESP and CAPES, Brazil.

\end{document}